# The Neuroscience of Spontaneous Thought: An Evolving, Interdisciplinary Field

Jessica R. Andrews-Hanna[1], Zachary C. Irving[2], Kieran C.R. Fox[3,4], R. Nathan Spreng[5], Kalina Christoff[3,6]

[1]Institute of Cognitive Science
University of Colorado Boulder
Boulder, CO, USA
Jessica.Andrews-Hanna@colorado.edu

[2]Department of Psychology
University of California Berkeley
Berkeley, CA, USA

[3]Department of Psychology
University of British Columbia
Vancouver, British Columbia, Canada

[4]Department of Neurology and Neurological Sciences
Stanford University Medical Center
Stanford, CA, USA

[5]Laboratory of Brain and Cognition
Human Neuroscience Institute
Department of Human Development
Cornell University
Ithaca, NY, USA

[6]Centre for Brain Health
University of British Columbia
Vancouver, British Columbia, Canada

**Abstract**

An often-overlooked characteristic of the human mind is its propensity to wander.  Despite growing interest in the science of mind-wandering, most studies operationalize mind-wandering by its task-unrelated *contents*, which may be orthogonal to the *processes* constraining how thoughts are evoked and unfold over time.  In this chapter, we emphasize the importance of incorporating such processes into current definitions of mind-wandering, and proposing that mind-wandering and other forms of spontaneous thought (such as dreaming and creativity) are mental states that arise and transition relatively freely due to an absence of constraints on cognition. We review existing psychological, philosophical and neuroscientific research on spontaneous thought through the lens of this framework, and call for additional research into the dynamic properties of the mind and brain.

## I. An introduction to an evolving, interdisciplinary field

A mere 10 years ago, the idea of an edited volume on spontaneous thought might have seemed far-fetched. Yet fast-forward to 2016, and the topic – once considered a "fringe" or "pseudo" science – has begun to thrive in mainstream research. This growing scientific interest in spontaneous mental activity was sparked by several independent findings from psychology and neuroscience research that have recently been synthesized under the heading of a new field: *the neuroscience of spontaneous thought* (see Christoff et al., 2016 for a recent review).

Beginning in the 1960s, findings from the psychology literature demonstrated that cognition often unfolds independent from the here-and-now (Singer & McCraven, 1961; Klinger & Cox, 1987; Kane, Brown & McVay, 2007; Killingsworth & Gilbert, 2010), and subsequent studies have shown that these *task-unrelated* or *stimulus-independent thoughts* exhibit complex relationships with attention (Antrobus, Singer & Greenberg, 1966; Teasdale et al., 1995; Smallwood & Schooler, 2006; McVay & Kane, 2010) and well-being (Giambra & Traynor, 1978; Watkins, 2008; McMillan, Kaufman & Singer, 2013). In parallel, neuroscientists discovered that a set of regions known as the *default network* becomes more active when participants disengage from a wide variety of tasks (Shulman et al., 1997a; Raichle et al., 2001), leading to a plethora of studies attempting to uncover the network's functional roles (reviewed in Buckner, Andrews-Hanna, & Schacter, 2008). Subsequently, the introduction of *resting state functional connectivity* (RSFC) into mainstream neuroscience research (i.e. Greicius et al., 2003; Fox et al., 2005) demonstrated that intricate maps of the brain's functional network architecture could be derived from an fMRI scan while individuals rested quietly in the scanner (reviewed in Fox & Raichle, 2007). Collectively, these findings revealed that a set of brain regions become engaged in coordinated ways while individuals are left alone undisturbed. Neuroscientists therefore started to question: "What is so special about periods of rest?"

In this chapter, we highlight how our understanding of the neuroscience of spontaneous thought has benefited greatly from integrating these parallel findings across psychological and neuroscientific levels of analysis, as well as related fields such as the philosophy of mind-wandering (Irving, 2016; Irving & Thompson, this volume; Carruthers, 2015; Dorsch, 2015; Metzinger, 2013; Metzinger, 2015; Sripada, this volume; Sripada, 2016). Realizing that the mind is always active – spontaneously associating, simulating, remembering, predicting, mentalizing, and evaluating – suggests that the default network's coordinated activity during periods of wakeful rest may neither be a coincidence nor indicative of a state of idleness. Similarly, the recent discovery that regions associated with executive control become engaged during mind-wandering[1] (Christoff et al 2009; Fox et al 2015) sheds important light on the complex behavioral relationships between mind-wandering and executive function. Here we discuss how interdisciplinary cross-talk led to evolving views on how to define, measure, and understand the significance of spontaneous thought, and how these inquiries continue to spark new questions for future research on this elusive phenomenon.

**II. Evolving definitions of spontaneous thought**

Although the phrase "spontaneous thought" is often equated with "mind-wandering" throughout the literature, we recently proposed that mind-wandering is but one member of a larger class of spontaneous processes that also includes nighttime dreaming, as well as aspects of creativity (Christoff et al., 2016; Figure 1). We defined spontaneous thought as "a mental state, or a sequence of mental states, that arise relatively freely due to an absence of strong constraints on the contents of each state and on the transitions from one mental state to another." Three key components of this definition are largely overlooked by prior research (see also Irving, 2016 for a philosophical theory that incorporates similar developments). First, the definition suggests that thoughts arising in a spontaneous or unintentional fashion should not be equated with thoughts arising deliberately, even when such thoughts have similar (e.g. task-unrelated) contents. Second, this definition contrasts thoughts that arise spontaneously from those that are constrained through automatic sources such as perceptual and affective salience. Third, if spontaneous thoughts unfold relatively free from constraints, they should flow in a flexible and dynamic manner.

--------------------------------

Insert Figure 1 here

--------------------------------

Although these principles may seem inherent to the term "spontaneity," the bulk of the mind-wandering literature characterizes the phenomenon by its *contents*, rather than the

---

[1] In Section 1, we discuss how the definition of mind-wandering varies throughout the literature, leading to disparate interpretations of existing experimental findings. Note that while we use the term "mind-wandering" loosely in this chapter, we are sensitive to these different interpretations and discuss them at length when possible.

*processes* by which thoughts are evoked (i.e. Smallwood & Schooler, 2006). For years, mind-wandering has been defined as being either unrelated to the task at hand (as a *task-unrelated thought*) (e.g. Giambra, 1989) or as independent from external stimuli (as a *stimulus-independent thought*) (e.g. Teasdale et al., 1995). While more recent taxonomies suggest that true episodes of mind-wandering are thoughts that are both task-unrelated *and* stimulus-independent (Stawarczyk et al., 2011a), such definitions do not consider the manner in which thoughts are evoked, nor how they unfold over time (but see Klinger, 1971; Irving, 2016; Irving & Thompson, this volume; Christoff, 2012; McMillan et al., 2013; Seli, Carriere, & Smilek, 2015a; Smallwood & Schooler, 2015; Stan & Christoff, this volume).

The distinction between spontaneous versus deliberate thought is critical in many respects. For one, recent research suggests that unintentional versus intentional task-unrelated thoughts show dissociable effects across a variety of behavioral and clinical contexts (Seli et al., 2016a). For example, intentional task-unrelated thoughts are most frequent in easy compared to difficult tasks, while unintentional thoughts are most frequent in difficult compared to easy tasks (Seli et al., 2016b). Further, greater endorsement of unintentional thinking, as measured with a trait questionnaire, positively predicts symptoms of both attention deficit-hyperactivity disorder (ADHD; Seli et al., 2015b) and obsessive-compulsive disorder (OCD; Seli et al., 2016c), despite the finding that intentional task-unrelated thoughts do not show significant relationships with symptoms of these disorders. Moreover, intentional task-unrelated thoughts positively predict aspects of trait mindfulness, while unintentional thoughts negatively predict the same mindfulness construct (Seli et al., 2015a). The distinction between unintentional and intentional task-unrelated thinking may also prove important when interpreting existing neuroscience research, as discussed in Section 4.

Another key dimension of spontaneous thought, foreshadowed by William James as the flowing "stream" in the *stream of consciousness* (James, 1890), is the manner in which thoughts unfold over time. According to the definition of Christoff and colleagues (2016), thoughts that concern a narrow topic, and remain fixated on this narrow topic over time, are not spontaneous in nature because of the excessive constraints that influence how one transitions from one thought to another. As discussed earlier, a train of thought can be constrained in two ways (Christoff et al., 2016; Irving, 2016). One type of constraint that can limit the flow of thought is deliberate in nature – i.e., evoked intentionally using top-down control, as when one chooses to remain focused on a particular topic for an extended period of time. Another type of constraint is automatic in nature – as when a habitual thought pattern or salient perceptual stimulus biases one's thoughts toward a specific topic in a bottom-up manner. This temporal variability, largely overlooked by prior research, has important clinical relevance. For example, excessive automatic constraints could characterize ruminative thoughts (DuPre & Spreng, this volume) – a common symptom of depression and anxiety (Nolen-Hoeksema, 2008; Watkins, 2008). In contrast, thoughts with excessive variability may characterize ADHD or aspects of psychosis (Christoff et al., 2016). The dynamics of spontaneous thoughts may have additional implications for recent neuroscientific findings (see Section 4).

Although this section has given much weight to *process* models of spontaneous thought, the *content* of spontaneous thought is also key, and variability in thought content over time is an important manifestation of its dynamic flow. Additionally, numerous studies have shown that task-unrelated thoughts can have a diverse array of content, including emotional, temporal, and social content that may differ within and between individuals in ways that relate to well-being (Singer & Antrobus, 1966; Klinger, 2009; Smallwood & Andrews-Hanna, 2013). For example, although meta-analyses of behavioral studies show that task-unrelated thoughts have a slightly positive bias on average (Fox et al., 2014; Fox et al., in preparation), symptoms of depression have been linked to more negative and self-focused thoughts (Giambra & Traynor, 1978; Andrews-Hanna et al., 2013). Additionally, while task-unrelated thoughts can sometimes predict worse subsequent mood (Killingsworth & Gilbert, 2010, but see Mason et al., 2013; Poerio et al., 2013), task-unrelated thoughts pertaining to the future predict *better* subsequent mood (Ruby et al., 2013). According to the *content regulation hypothesis* proposed by Smallwood and Andrews-Hanna (Smallwood & Andrews-Hanna, 2013; Andrews-Hanna, Smallwood, & Spreng, 2014), an ability to limit one's task-unrelated thoughts to largely positive, constructive content is thought to be a critical factor governing its costs and benefits.

**III. Evolving approaches to measuring the neuroscience of spontaneous thought**

Thus far, this chapter has introduced a new field – the neuroscience of spontaneous thought – and discussed how the definition of spontaneous thought (and mind-wandering, in particular) has evolved in recent years. Before synthesizing findings from research on this topic, it is worth discussing how the neural underpinnings of spontaneous thought are commonly measured. The element of spontaneity poses a unique challenge inherent to its experimental study. How can one measure a process that, by definition, cannot be directly experimentally induced, as doing so would introduce deliberate constraints on cognition that conflict with spontaneity? And how can one isolate stretches of spontaneous thought, when they arise at unpredictable times, independently of immediate perceptual input and experimental demands, and often unbeknownst to the person having those thoughts? The next section reviews evolving approaches to measuring the neural underpinnings of spontaneous thought, and evaluates such approaches in light of the definitions discussed in Section 2.

*Early neuroscience research measured spontaneous thoughts accidentally and indirectly*

Although the field of psychology had begun to address the challenges inherent to the measurement of spontaneous thought by the 1990s, historical biases and demands for rigorous experimental control pressured the neuroscience field to focus on externally-oriented processes with measurable behavioral manifestations (Callard, Smallwood & Margulies, 2012). As a result, the neuroscience of spontaneous thought trailed behind for decades (but see early efforts by Ingvar et al., 1985; Andreasen et al., 1995; McGuire et al., 1996; Binder et al., 1999). Given these biases, it may not seem surprising that the *default network,* a brain system now widely appreciated for its role in internally-directed thought (Buckner, Andrews-Hanna, Schacter, 2008), was discovered entirely accidentally. This ground-breaking discovery followed a meta-analysis of nine different Positron Emission Tomography (PET) studies of "human visual information processing," each with passive control conditions in which participants fixated on a crosshair or passively viewed the same stimuli (Shulman et al. 1997a; 1997b). To the surprise of the researchers, relatively few regions would exhibit common patterns of blood flow increases across the experimental tasks (Shulman et al., 1997b), while a robust set of regions would show the opposite contrast of *passive fixation > active tasks* (Shulman et al., 1997a). The network that emerged in this second comparison was coined the "default mode of brain function" by Raichle and colleagues in 2001 (Raichle et al., 2001).

Although these two manuscripts brought initial attention to the default mode and introduced several hypotheses regarding the default network's functional significance during periods of awake rest, including the generation of "unconstrained verbally-mediated thoughts" (Shulman et al., 1997a), the studies did not assess the frequency or nature of ongoing thoughts during periods of rest. Despite the efforts of these groups, many subsequent studies assumed the default network and the resting state reflected an idle state with little contribution to active forms of cognition. This assumption was perhaps most apparent throughout the literature on rs-fcMRI, a technique that examines temporally correlated fMRI activity patterns during extended periods of awake rest (reviewed in Fox & Raichle, 2007). In 2003, fMRI activity time courses from key regions of the default mode were shown to temporally correlate at low frequencies during the resting state, forming a brain system known as the *default mode network*, or *default network* (Grecius et al., 2003; 2004). Several other large-scale brain systems have been subsequently identified using principles of rs-fcMRI (Yeo et al., 2010; Power et al., 2010; Doucet et al., 2011). A commonly held assumption of rs-fcMRI was that patterns of connectivity are *intrinsic* in nature, reflecting a long history of firing and wiring (Fox & Raichle, 2007). Periods of awake rest were used to evoke resting state correlations because cognition was assumed to be at a minimum during this unconstrained state, and low temporal frequencies were isolated partially to ensure that task-related activity was filtered out, despite later findings that unconstrained thoughts unfold at similar frequencies (Klinger, 2009; Vanhaudenhuyse et al., 2010).

Thus, ground breaking discoveries from neuroscience research in the early 2000s revealed that periods of awake rest were associated with increased activity in a set of regions that came to be known as the default mode network. Scientists became curious about periods of rest, prompting a synthesis of the psychological literature on unconstrained cognition and mind-wandering. These initial efforts revealed that the absence of experimental tasks should not be equated with the absence of cognition (Andreasen et al., 1995; Binder et al., 1999; Christoff, Ream & Gabrieli, 2004; Buckner, Andrews-Hanna, & Schacter, 2008), and set the stage for an explosion of research to come.

*Measuring the neuroscience of spontaneous thought in the modern age*

The appreciation that thoughts frequently unfold in the absence of internal and external constraints on cognition led to a plethora of neuroimaging and electrophysiological studies attempting to more precisely characterize their neural underpinnings. Here we review mainstream methods to measure the neuroscience of spontaneous thought (Figure 2).

-------------------------------
Insert Figure 2 here
-------------------------------

One common method examines individual difference relationships between covert neurocognitive measures (such as fMRI activity, strength of rs-fcMRI correlations between brain regions, structural MRI, neurophysiological / occulometric measures, electroencephalography (EEG) and event related potentials (ERP)) and participant scores on *trait questionnaires* assessing the typical nature of spontaneous thoughts in daily life. Examples of such questionnaires include the Imaginal Process Inventory (Singer & Antrobus, 1966), the Mind-Wandering Questionnaire (Mrazek et al., 2013), and the recent Mind Excessively Wandering Scale (Mowlem et al., 2016). Scores on these and other trait questionnaires are correlated across participants with individual differences in brain activity or connectivity during experimental tasks (Mason et al., 2007) or periods of rest (Kucyi & Davis, 2014). Of particular interest given evolving definitions of spontaneous thought are two additional scales that separately assess the tendency for individuals to engage in intentional and unintentional forms of thought: the Mind-Wandering Deliberate Scale and the Mind-Wandering Spontaneous Scale (Carriere et al., 2013; Seli, Carriere, & Smilek, 2015a). Additionally, the "Mentation Rate" and "Absorption in Daydreams" subscales of the Imaginal Process Inventory (Singer & Antrobus, 1966) focus on the dynamics of spontaneous and deliberate thought (e.g. how quickly one's thoughts transition from topic to topic versus how likely thoughts are to remain focused on a specific topic). Several clinically-oriented questionnaires may indirectly measure the dynamics of thought, insofar as they plausibly measure automatic constraints on thought processes. These inventories include the Ruminative Response Scale (Roberts, Gilboa & Gotlib, 1998; Treynor, Gonzalez, & Nolen-Hoeksema, 2003), the Rumination-Reflection Questionnaire (Trapnell & Campbell, 1999), the Cognitive Intrusions Questionnaire (Freeston, 1991), and the Intrusive Thoughts Questionnaires (Edwards & Dickerson, 1987). Recent studies have begun to examine relationships between these clinically-focused traits and individual differences in brain activity or connectivity (Hamilton et al., 2011; Kaiser et al., 2015; Ordaz et al., 2016).

Relating neurobiological measures to trait questionnaires has the advantage of allowing researchers to assess more stable properties of mind-wandering, but may not provide an accurate assessment of participants' thoughts during the tasks or rest periods for which neurobiological measures are derived. For example, Berman et al. (2014) found that group differences in functional connectivity between depressed and non-depressed individuals were much more substantial following a rumination induction period than during a baseline resting state, suggesting that spontaneous cognition may not always track trait measures. To overcome this limitation, researchers commonly administer *retrospective questionnaires* after task paradigms or periods of rest in which neurobiological measures are simultaneously recorded. Retrospective questionnaires require participants to retrospectively reflect on their phenomenological experience during those paradigms, and answer a series of self-report questions characterizing the nature of their thoughts during that time. This approach was originally implemented in conjunction with neuroimaging on an informal verbal basis, prompting Andreasen and colleagues to coin the ironic acronym *Random Episodic Silent Thought* to emphasize that periods of rest often involve autobiographical memory recall and future thought (Andreasen et al., 1995). In 1996, McGuire and colleagues (McGuire et al., 1996) used a retrospective questionnaire to assess participants' frequency of task-unrelated thoughts following task and rest conditions, and examined which brain regions tracked individual differences in the frequency of task-unrelated thoughts (see also Andrews-Hanna et al., 2010a). More than a decade later, a retrospective Resting State Questionnaire was developed to quantify the content and form of thoughts during resting state scans (Delamillieure et al., 2010), and a similar questionnaire was administered in a different study that linked individual differences in thought content to individual differences in functional connectivity during rest (Andrews-Hanna et al., 2010a; see also Doucet et al., 2012; Gorgolewski et al., 2014). Across these studies, participants reported spending a large proportion of time engaging in stimulus-independent thoughts, with thoughts about the future being especially frequent. Retrospective questionnaires have also been used in conjunction with methods assessing pupillometry (Smallwood et al., 2012) and EEG (Barron et al., 2011). Collectively, retrospective questionnaires have the potential to reveal the frequency and content of spontaneous thoughts without disrupting ongoing cognition or biasing subsequent attention. A drawback of this approach is that participants may not remember the contents of their thoughts when assessed minutes later, and reported thoughts might be influenced by biases in memory. Additionally, since retrospective questionnaires often ask participants to average multiple thoughts across extended periods of time, they are not ideal approaches to examine the way in which thoughts precisely unfold over time.

*Online experience sampling approaches* have the potential to overcome many of the limitations of retrospective questionnaires by assessing the nature of thoughts at different moments throughout a task. This approach had gained popularity in psychology (Giambra, 1989; Teasdale et al., 1995), and had initially been adopted for use in simulated scanning environments to approximate the frequency of task-unrelated thoughts in identical paradigms conducted in the scanner (Binder et al., 1999; McKiernan et al., 2006; Andrews-Hanna et al., 2010a). It was not until 2009, however, that experience sampling probes were incorporated directly into neuroimaging paradigms, allowing researchers to compare patterns of brain activity associated with epochs of on-task versus off-task thought, and examine correspondences with disruptions in behavioral performance (Christoff et al., 2009a). Since then, additional studies have adopted similar online experience sampling approaches during periods of rest (Vanhaudenhuyse et al., 2010; Tusche et al., 2014; Van Calster et al., 2016) or external tasks (Stawarczyk et al., 2011b; Kucyi, Salomons & Davis, 2013), and the approach has also gained popularity in the EEG (Smallwood et al., 2008; Kam et al., 2011; Kirschner et al., 2012; Kam et al., 2013; Baird et al., 2014), and structural MRI literature (Bernhardt et al., 2014). Note, however, that while most studies ask participants whether they characterized their thoughts as on-task or off-task, some studies have additionally asked about participants' meta-awareness of their thoughts (i.e. Christoff et al., 2009a), or whether their thoughts were dependent or independent of external stimuli (Stawarczyk et al., 2011b; Kucyi, Salomons & Davis, 2014). These questions, as well as questions assessing other measures of phenomenological content, are important because participants are often unaware of their ongoing mental activity (Schooler et al., 2011; Fox & Christoff, 2015). Furthermore, a sizeable proportion of off-task thoughts pertain to external distractions and involve unique neural underpinnings (Stawarczyk et al., 2011b). Critically, to our knowledge, no neuroscientific study has directly assessed using online experience sampling approaches whether participants' thoughts arose in a spontaneous or constrained fashion. Thus, as discussed in the next section, existing neuroimaging research may present an incomplete picture, and resolving the neuroscience of spontaneous versus deliberate (and automatically constrained) thought marks an important direction for future research.

One twist on the online experience sampling approach allows participants to press a button the moment they become aware of a spontaneous thought arising. Participants then answer questions characterizing their thought, and subsequently return their attention back to the ongoing task (e.g., focusing on one's breath). This *self-caught online thought sampling* approach gives researchers the opportunity to analyze patterns of brain activity before, during, and after moments of awareness of thought, differentiating brain regions involved in the generation and awareness of task-unrelated and/or spontaneous thoughts (Ellamil et al., 2016; see also Hasenkamp et al., 2012 and Hasenkamp, this volume). Related approaches have been adopted for non-meditators as well (Vanhaudenhuyse et al., 2010), offering additional insight into the dynamics of thinking.

**IV. Evolving insight into the neurobiological basis of spontaneous thought**

As discussed in the last section, neurocognitive research over the last several years has witnessed evolving methodological approaches to characterizing brain systems and tracking the frequency and phenomenology of spontaneous thought. Indirect or inferential approaches have begun to be complemented by real-time assessments that are less influenced by biases and failures of memory. Avenues for future research include novel approaches that could assess spontaneous thoughts covertly without interrupting the participant or relying on self-report, as well as methods that could examine the causal role of brain regions in the generation or dynamics of spontaneous thought (see Section 5). In this section, we ask what can be gleaned from the research outlined above on the neuroscience of spontaneous thought (see also Kam & Handy, this volume, for a summary of ERP research).

*The role of the brain's default network in mind-wandering[2]*

As discussed above, early neuroimaging studies observed that short breaks in between blocks of externally-directed tasks led to blood flow increases in brain regions that would be come to be known as the default network. Subsequent rs-fcMRI studies determined that activity fluctuations within these structures are temporally correlated during rest (i.e. Grecius et al., 2003; Fox et al., 2005). Clustering the magnitude of interregional associations during rest and tasks revealed that the default network can be further parceled into two *subsystems* that converge on *core hubs* (Andrews-Hanna et al., 2010b; Yeo et al., 2010). A ventrally-positioned Medial

---

[2] Strictly speaking, we have argued that mind-wandering is a form of spontaneous thought, and so we cannot be sure that these studies are measuring mind-wandering properly defined (as we discuss throughout this section when interpreting current findings).

Temporal subsystem includes the hippocampal formation and parahippocampal cortex, as well as two cortical regions that exhibit direct anatomical connections with the medial temporal lobe: ventral angular gyrus and retrosplenial cortex. A Dorsal Medial subsystem includes structures spanning the dorsal medial prefrontal cortex (mPFC), the temporoparietal junction, the lateral superior and inferior prefrontal gyrus, and the middle temporal gyrus extending into the temporal pole. These subsystems are strongly interconnected with a set of core hubs centered on the anterior mPFC, the posterior cingulate cortex, the dorsal angular gyrus, the superior frontal sulcus, and right anterior temporal cortex. The default network also includes aspects of Crus 1 and II of the cerebellum (Buckner et al., 2014), and subcortical regions such as aspects of the dorsal and ventral striatum (Choi et al., 2014). Thus, structural and functional MRI research suggests the default network is a large brain system with interacting components that converge on key association cortices.

Although early neuroimaging studies did not explicitly assess participants' mental states during periods of "rest" that give rise to default network activity, links between unconstrained thinking and the default network were observed across several subsequent studies employing a variety of methodological approaches (but see, Raichle 2016). For example, retrospective questionnaires and online experience sampling approaches revealed that conditions in which participants reported high frequencies of task-unrelated thought were associated with greater default network activity (Mason et al., 2007; McKieran et al., 2006; Andrews-Hanna et al., 2010a), that off-task trials activated the default network to a greater degree than on-task trials (Christoff et al., 2009a; Stawarczyk et al., 2011b), and that stimulus-independent thoughts during the resting state activated the default network to a greater degree than epochs in which participants were focused on external perceptions (Vanhaudenhuyse et al., 2010; Preminger, Harmelech, & Malach, 2011; Stawarczyk et al., 2011b; Van Calster et al., 2016). Additionally, individual difference analyses revealed positive associations between default network activity during ongoing tasks and mind-wandering as assessed with trait questionnaires and retrospective measures (Mason et al., 2007). Conversely, experienced meditators, as compared to novices, show less default network activity and experience fewer task-unrelated thoughts while meditating (Brewer et al., 2011). Links between task-unrelated thinking and the default network also extend to individual differences in rs-fcMRI measures (Wang et al., 2009; Andrews-Hanna et al., 2010a; Doucet et al., 2012; Gorgowleski et al., 2014; O'Callaghan et al., 2015; Smallwood et al., 2016), and structural MRI measures such as cortical thickness (Bernhardt et al., 2014; Golchert et al., 2016). In 2015, much of the work outlined above was synthesized in two formal fMRI meta-analyses of mind-wandering (Fox et al., 2015; Stawarczyk & D'Argembeau, 2015). These meta-analyses revealed that several regions throughout the default network – particularly regions associated with the DN core – were reliably associated with mind-wandering across studies employing diverse populations and methodological approaches (Figure 3), but also regions outside the DN (see below).

-------------------------------
Insert Figure 3 here
-------------------------------

Although the default network is now widely appreciated for its role in spontaneous thought, the precise functional contributions of the specific regions involved remain unclear, particularly because the studies included in the meta-analysis defined mind-wandering by its task-unrelated and/or stimulus-independent nature rather than by its spontaneous dynamic processes. One intriguing possibility is that different regions, subsystems, or multivariate patterns within the default network support the conceptual content and/or form characterizing spontaneous thoughts (Andrews-Hanna et al., 2010; Andrews-Hanna, Smallwood & Spreng, 2014; Tusche et al., 2014; Gorgolweski et al., 2014; Smallwood et al., 2016). For instance, specific patterns of default network activity might differentiate a positive spontaneous thought about one's upcoming wedding from a negative memory about an ex-partner, or a thought of any other conceptual nature. This possibility is supported by evidence from a variety of task-related and rs-fcMRI studies suggesting that the Medial Temporal subsystem might support contextual, visuospatial and temporal aspects of memory and imagination – important for constructing a mental scene (Bar, 2007; Bar et al., 2007; Hassabis & Maguire, 2009; Addis et al., 2009; Andrews-Hanna et al., 2010b) – whereas the Dorsal Medial subsystem may support a variety of socio-emotional content (Lieberman, 2007; Andrews-Hanna, Saxe & Yarkoni, 2014; Spreng & Andrews-Hanna, 2014; Hyatt et al., 2015). The widespread connectivity of core hub regions, combined with their involvement in a variety of self-generated processes, well-position these regions to integrate disparate conceptual information when computing the overarching significance or importance of a particular thought (Andrews-Hanna, Smallwood, & Spreng, 2014; Smallwood et al., 2016) – a process that may partly determine the thought's dynamics, or the way in which it unfolds overtime.

It is also possible that specific components of the default network directly contribute to the generation of spontaneous thoughts in a more domain-general manner, yet interact with other default regions, such as the lateral temporal cortex, to elaborate thoughts with specific conceptual content (e.g. Szpunar et al., 2015). Evidence from human neuroimaging, intracranial recordings, rodent neurophysiology, and lesion work suggests that the hippocampus and nearby medial temporal structures may be prime candidates for such components (for reviews, see Fox, Andrews-Hanna & Christoff, 2016; Christoff et al., 2016). In particular, activity in the hippocampus, parahippocampal cortex, and other aspects of the Medial Temporal subsystem emerge early in the dynamics of spontaneous thought – just prior to the moment of subjective awareness – consistent with a role in the initiation as opposed to the elaboration and/or evaluation of spontaneous thought (Gelbard-Sagiv et al., 2008; Ellamil et al., 2016). Additionally, reviews of human intracranial recording studies suggest that spontaneous thoughts, memories, and other dreamlike experiences are elicited more than half the time following electrical stimulation of regions within the medial temporal lobe – considerably more than any other cortical region assessed (reviewed in Selimbeyoglu and Parvizi, 2010; Fox et al., 2016). In rats, hippocampal place cells – neurons with spatial receptive fields that track where a rat is in its environment (O'Keefe & Nadel, 1978) – also spontaneously fire independent of immediate perceptual input, including during brief epochs of "rest" when the rat stops navigating its environment (Foster & Wilson, 2006). This spontaneous hippocampal firing has been linked to replay of prior experiences (Foster & Wilson, 2006; Diba and Buzsaki, 2007), pre-play of upcoming experiences (Dragoi & Tonegawa, 2011, 2013), and even to patterns suggestive of simulations of entirely novel experiences (Gupta et al., 2010). In humans, spontaneous hippocampal activity and connectivity during periods of rest following periods of learning predict the degree to which studied material is encoded into long-term memory (Tambini, Ketz, & Davachi, 2010), and periods of rest as well as sleep are considered critical for memory consolidation and problem solving (Wagner et al., 2004; Stickgold, 2005; Dewar et al., 2012, 2014). Finally, damage to the medial temporal lobe in hippocampal amnesia and Alzheimer's disease is associated with profound deficits in both memory and imagination (Hassabis et al., 2007; Irish & Piolino, 2016), although the effect of such lesions on spontaneous thought has yet to be investigated to our knowledge. In sum, multiple sources of evidence from human and non-human animals suggest that the medial temporal lobe may play a key role in the initiation of a spontaneous thought. The medial temporal lobe is densely interconnected with cortical structures throughout the Medial Temporal subsystem as well as several core default network regions such as the lateral temporal cortex (Suzuki & Amaral, 1994; Lavenex & Amaral, 2000), thought to play an important role in conceptual knowledge and elaboration (Patterson, Nestor, & Rogers, 2007; Rice, Lambon Ralph & Hoffman, 2015). Thus, it is likely that the medial temporal lobe does not operate in isolation, and we suspect its connectivity with distant cortical regions within and outside the default network are important determinants of the phenomenological content, form, dynamics, and conscious awareness of spontaneous thought.

*The role of the frontoparietal control network in mind-wandering*

Although the role of the default network in spontaneous thought has now gained support from a considerable body of research, neuroimaging meta-analyses of mind-wandering also reveal reliable involvement of several regions *outside* the default network (Figure 3) (Fox et al., 2015; but see Stawarczyk & D'Argembeau, 2015). Most notable are aspects of the frontoparietal control network (FPCN), a set of regions spanning association cortices such as the lateral prefrontal cortex, dorsal anterior cingulate / pre-supplementary motor area, and anterior inferior parietal lobe (Vincent et al., 2008; Yeo et al., 2010). The FPCN is thought to allow individuals to flexibly allocate attentional resources towards external stimuli and/or internal representations (i.e. thoughts, memories, and emotions), and integrate relevant information from external and internal sources of information in the service of immediate and long-term goals (Vincent et al., 2008; Cole et al., 2013; Spreng et al., 2013). While the involvement of the FPCN in task-unrelated and/or stimulus-independent thought may seem surprising given that mind-wandering is often assumed to reflect a failure of control (McVay & Kane, 2010; Kane & McVay, 2012), a closer look at the data points to many possible explanations that mark important avenues for future research.

First, the majority of studies define mind-wandering by its task-unrelated and/or stimulus-independent contents, and consequently lump spontaneous and deliberate thoughts together when conducting analyses. It is therefore possible that the FPCN comes online only

To appear in: *The Oxford Handbook of Spontaneous Thought* (Fox & Christoff, Eds.)     16

when individuals experience deliberate task-unrelated thoughts. There are two ways to interpret this claim, which arise from the two ways to distinguish spontaneous from deliberate thoughts. First, it is possible that the FPCN comes online only when participants deliberately disengage from ongoing tasks, and that thoughts arising without intention may not involve the FPCN at all (Seli et al., 2016b). Second, it is possible that the FPCN comes online only when participants deliberately constrain the course that their thoughts take as they unfold over time. In this sense, transient FPCN activity may reflect the deliberate re-allocation of attention away from the task at hand and/or the sustained pursuit of internal goals that are irrelevant to the task at hand. Similarly, sustained patterns of FPCN activity may help participants shield their internal thoughts from less personally-significant distractions, including the sounds of the scanning environment, other thoughts deemed less important, or even the task itself. Despite the injunctions to participants to stay alert and focused on external stimuli in paradigms, such as the Sustained Attention to Response Task (SART), participants may consider real-world issues such as an upcoming exam, a weekend trip, or an unresolved conflict with a friend more pressing "tasks," which may therefore vie for attention in potentially adaptive ways (Baars, 2010; Andrews-Hanna, 2012; McMillan et al., 2013). Nevertheless, recent findings provide support for the idea that the FPCN might play an important role in deliberate (but not necessarily spontaneous) task-unrelated thinking. Golchert and colleagues (2016) used a trait questionnaire to assess the frequency with which participants engaged in deliberate and spontaneous forms of task-unrelated thought. Participants who engaged in more frequent deliberate thinking exhibited greater functional integration between the FPCN and the DN, and greater cortical thickness in aspects of the FPCN.

   Another explanation for the FPCN's involvement in spontaneous and/or deliberate task-unrelated thought concerns the phenomenological content characterizing periods of spontaneous or deliberate thought. Behavioral and neuroimaging studies exploring the content of task-unrelated thoughts suggest that adults spend a considerable proportion of time engaged in prospectively-oriented thoughts, including thoughts about future goals, and in planning how to achieve those goals (Andrews-Hanna et al., 2010a; Baird, Smallwood & Schooler, 2011; Song & Wang, 2012; Andrews-Hanna et al., 2013; Stawarczyk et al. 2011a, 2013). Support for this idea comes from neuroimaging studies using task paradigms in which participants are explicitly asked to plan for their future. In these autobiographical planning contexts, both the default network and FPCN become engaged in a tightly coordinated manner (Spreng et al., 2010; Gerlach et al., 2011; Gerlach et al., 2016; Spreng et al., 2015). One intriguing question is whether thoughts pertaining to upcoming goals can occur and/or unfold in a spontaneous fashion, as suggested by Klinger's *current concern hypothesis* (Klinger, 1971; Klinger, 2009). If so, would such thoughts recruit activity within the FPCN?

The nature of the task paradigms employed in mind-wandering studies may also partly explain why activity within the FPCN is often associated with task-unrelated thoughts. Most neuroscience research assesses mind-wandering retrospectively or with online experience sampling probes during easy behavioral paradigms in which participants can maintain a minimal level of performance while *simultaneously* allocating their attention towards the task at hand and their (potentially) spontaneous thoughts. Thus, easy tasks – including resting state paradigms and the SART task – may encourage dual-task situations where participants simultaneously direct their attention externally and internally, or rapidly switch between internal and external modes of attention. Both types of processes may recruit the FPCN in a regulatory manner to help coordinate attention across tasks. Conversely, maintaining reasonable levels of performance during difficult behavioral paradigms may require that participants be continuously focused on the task at hand. In these scenarios, occasional off-task trials may be more likely to manifest as decreases in FPCN activity, reflecting lapses in attention marked by a failure of control. Paralleling these findings, behavioral studies of mind-wandering reveal complex relationships between mind-wandering and executive function that appear to partly depend on the difficulty of the ongoing task (Smallwood & Andrews-Hanna, 2013). For example, participants who frequently experience task-unrelated thoughts during easy tasks have higher working memory capacity (Levinson, Smallwood & Davidson, 2013), suggesting that they are simultaneously able to have task-unrelated thoughts while maintaining acceptable performance on the task. Conversely, participants who frequently experience task-unrelated thoughts during difficult tasks tend to exhibit poorer working memory capacity (Kane & McVay, 2012; Unsworth & McMillan, 2013). These findings prompted Smallwood and Andrews-Hanna to propose the *context regulation hypothesis*, suggesting that the costs and benefits of mind-wandering partly depend on an individual's ability to constrain task-unrelated thoughts to easy or unimportant contexts (Smallwood & Andrews-Hanna, 2013; Andrews-Hanna, Smallwood & Spreng, 2014). In light of these observations, future studies could consider examining patterns of activity associated with task-unrelated thoughts across tasks that vary in difficulty.

Finally, it is important to note that not all regions within the FPCN are reliably engaged across studies of mind-wandering: whereas rostral lateral PFC (rlPFC) and dACC/pre-SMA are among those present, meta-analyses of mind-wandering show that dlPFC, posterior PFC and anterior inferior parietal lobe are not reliably engaged across studies (Fox et al., 2015). The rlPFC has been linked to metacognitive awareness of one's thoughts, attention and performance (McCaig et al., 2011; Fleming & Dolan, 2012), so activity in the rlPFC could reflect the monitoring processes encouraged by online experience sampling tasks (but see Christoff et al., 2009a). Furthermore, dACC/pre-SMA may play an important role in the detection of internal conflict elicited when mind-wandering during the presence of an ongoing task, or in computing tradeoffs in the expected values of being on-task versus off-task.

Intriguingly, according to some theories of prefrontal cortex function, the prefrontal cortex is organized along a rostro-caudal gradient, where more anterior regions become engaged by more abstract or temporally-extended conditions that often rely on internal processes such as episodic memory and maintenance of long-term goals (Christoff & Gabrieli, 2000; Christoff et al., 2003; Badre & D'Esposito, 2009; Christoff et al., 2009b; O'Reilly, 2010; Dixon, Fox & Christoff, 2014). Conversely, posterior PFC regions tend to respond to more specific task demands, such as learning specific stimulus-response contingencies, biasing attention towards one stimulus attribute and away from another, or flexibly adjusting one's attention following errors in performance (Badre & D'Esposito, 2009; Christoff et al., 2009b; O'Reilly, 2010). It is therefore possible that the FPCN activation during mind-wandering represents a form of abstract control that is compatible with a considerable degree of dynamic spontaneity. Because abstract goals (e.g. "do well as an academic") place few constraints on *how* one is to achieve or think about them, one's thoughts may be directed towards such a goal, while still spontaneously wandering to a broad range of ideas (Irving, 2016). This hypothesis makes testable predictions: periods of spontaneous task-unrelated thought (measured through online thought sampling) that are loosely constrained to an abstract goal should be associated with more anterior PFC activation, whereas periods of task-unrelated thought that are deliberately constrained to a specific goal should be associated with more posterior PFC activation. In sum, although the functional contributions of the FPCN are unclear, different regions or patterns within and across regions likely play different roles and are likely influenced by a variety of factors (see Christoff et al., 2016 for a similar point).

*The role of additional brain regions*

A synthesis of existing neuroscience research on mind-wandering would be incomplete without discussing the involvement of additional regions outside the default and frontoparietal networks, including the lingual gyrus, somatosensory cortex, and posterior insula (Figure 3). Collectively, these regions may be associated with the sensory and embodied perception of task-unrelated thoughts. Many individuals characterize their thoughts during awake rest as unfolding in the form of mental *images* (Delamillieure et al., 2010), and the lingual gyrus may support the visual nature of such thoughts. The lingual gyrus is reliably observed during a variety of visual and mental imagery tasks – including when individuals are dreaming (Fox et al., 2013), or asked to recall their past and imagine their future (Addis et al., 2009) – and lesions to this region are associated with an impaired ability to engage in visual imagery and reduced levels of dreaming (Solms et al., 1997, 2000). Similarly, the somatosensory cortex and posterior insula, associated with tactile sensation, tactile imagery, and interoceptive awareness (Craig, 2003; Seung-Schik et al., 2003; Critchley et al., 2004), could relate to the frequently-reported thoughts about the body (Delamillieure et al., 2010; Diaz et al. 2013) and/or distracting external sensations. Interestingly, there is some evidence from rs-fcMRI that individuals who characterize their thoughts as having more visual imagery during periods of rest show heightened connectivity between visual regions, including the lingual gyrus, somatosensory cortices, and posterior insula (Doucet et al., 2012), perhaps reflecting attention towards sensorimotor and perceptual characteristics of unconstrained thinking.

*The role of default and frontoparietal networks in dreaming and creative thought*

In this section, we have synthesized research investigating the neuroscience of spontaneous thought, focusing on studies defining mind-wandering largely by its contents (since studies defining mind-wandering by its spontaneity are scarce). Whereas some expected findings have emerged from this synthesis – namely, the involvement of the default network – other findings – namely, the involvement of the FPCN – are more surprising, inviting several distinct hypotheses regarding their precise role in spontaneous thought. We now turn to neuroscience research examining two cognitive processes closely related to spontaneous thought – namely, dreaming and creative thinking – and ask whether our knowledge of these mental states can shed light on the role of the FPCN in mind-wandering and spontaneous thought more broadly.

According to the dynamic process model of spontaneous thought illustrated in Figure 1, dreaming is considered more spontaneous than mind-wandering due to an absence of many constraints on the contents and flow of mental states during this period, resulting in the bizarre, improbable, and highly dynamic characteristics of dreams (Hobson et al., 2000). Conversely, creative thought is considered less spontaneous than mind-wandering because it usually unfolds in the service of a specific goal (e.g., to generate a creative idea, solution, or product), and involves more deliberate processes of selecting a creative solution, evaluating its utility, and revising it if necessary (Beaty et al., 2015a). Creative insight also involves aspects of metacognitive awareness (Armbruster, 1989), as individuals who lack awareness of their creative ideas may be unable to benefit from them. Consequently, assuming that activity within the FPCN during mind-wandering at least partially reflects the deliberate nature of task-unrelated thoughts, and/or the metacognitive awareness that often accompanies them, one might expect that dreams would show activity reductions in the FPCN (consistent with lack of cognitive control and metacognitive awareness), while creative thinking might evoke increases in FPCN activity, particularly during later evaluative stages of the creative process.

A synthesis of neuroimaging literature on dreaming supports the role of the FPCN in deliberate constraints and metacognitive awareness (Fox et al., 2013). Compared to periods of relaxed wakefulness, REM sleep is associated with enhanced activity throughout the default network's Medial Temporal subsystem, and reductions in activity throughout the FPCN, consistent with the bizarre nature of dreams and a lack of awareness while dreaming (Figure 4a). Interestingly, lucid dreamers – individuals who are aware of their dreams while dreaming, and are often able to deliberately control how their dreams unfold – have enhanced grey matter volume throughout rostrolateral and medial PFC (Filevich et al., 2015), and also exhibit enhanced rlPFC activity during tasks in which participants are explicitly asked to monitor the contents of their thoughts (Filevich et al., 2015). The FPCN is also widely recruited during lucid REM sleep as compared to non-lucid REM sleep (Dresler et al., 2012).

-------------------------------
Insert Figure 4 here
-------------------------------

Conversely, many creative tasks, including tasks of divergent thinking, poetry generation, and creative idea generation, show initial activity within the Medial Temporal subsystem and the posterior cingulate cortex, followed by enhanced activity and connectivity of FPCN regions when deliberate constraints must be implemented to hone in on creative ideas, or evaluate and revise creative products (Figure 4b) (Ellamil et al., 2012; Liu et al., 2015; Beaty et al., 2015ab). Thus, evidence from neuroscience research on dreaming and creative thinking suggests that the involvement of the FPCN in mind-wandering – and spontaneous forms of thinking more broadly – might reflect deliberate control processes that serve to constrain the content and flow of mental states by guiding and suppressing their spontaneity (Fox & Christoff, 2014). In short, the growing body of neuroscience findings on mind-wandering may reflect an intricate balance of spontaneous and deliberate cognitive processes, and it remains a task for future research to unravel the common or distinct neural underpinnings of each.

**V. New promises for future inquiry into the neuroscience of spontaneous thought**

In this chapter, we have synthesized an interdisciplinary field of inquiry – the *neuroscience of spontaneous thought* – and discussed how definitions of spontaneous thought, approaches to measure spontaneous thought, and knowledge of brain systems supporting such thoughts have rapidly evolved in a few short years. New definitions rely less on the task-unrelated and stimulus-independent content that had long dominated the literature, and more on the processes that govern their initiation, as well as the temporal dynamics that characterize their flow; measurement approaches have shifted from indirect to more direct approaches in which thoughts are assessed much closer to the time at which they occur; and recent neuroscience findings emphasize the importance of regions outside the default network – such as the FPCN, sensorimotor networks, and posterior insula.

With exciting theoretical and methodological progress come many new questions and avenues for future research. One timely research direction regards assessing the forgotten dynamics of spontaneous thought (Christoff et al., 2016; Irving, 2016). To this end, online experience sampling approaches could be used in conjunction with dynamic rs-fcMRI (Calhoun et al. 2014) to elucidate how spontaneous thoughts and their corresponding neural underpinnings unfold and change over time (Zabelina & Andrews-Hanna, 2016). These approaches may also shed light on key mechanisms underlying a variety of mental health disorders (e.g., Kaiser et al., 2015).

Another direction for future research includes differentiating between spontaneous and deliberate thoughts at both the behavioral and neural level, and resynthesizing existing research in light of subsequent findings. Relatedly, there is some suggestion that spontaneous thoughts can be further characterized by the unintentional manner in which they are *initiated* (i.e. Seli et al., 2016b), as well as the unconstrained nature in which they *unfold* over time (Christoff et al., 2016; Irving, 2016; see also Stan & Christoff, this volume). Although the initiation and dynamics of spontaneous thought are likely correlated, they are conceptually distinct (Smallwood, 2013). They are likely correlated because top-down constraints on thought are typically (perhaps always) initiated with deliberate intent. Yet they are conceptually distinct for two reasons. For one, it seems possible to unintentionally initiate automatic constraints on the spontaneous dynamics of thought. For example, a depressed patient might unintentionally begin to ruminate. Furthermore, it seems possible to intentionally initiate a thought process with spontaneous dynamics. During a boring lecture, for example, one might intentionally let one's mind wander in an unconstrained manner. To help determine the relationship between these two ways of characterizing spontaneous thought, future research could design questionnaires that directly assess the tendency to have thoughts whose *dynamics* are unconstrained, compared to the tendency to *initiate* thoughts unintentionally. Researchers could then relate both forms of spontaneity to individual differences in brain activity and connectivity.

Additionally, although this chapter has largely focused on the role of deliberate constraints in restricting the contents and flow of thought, another type of constraint can also limit its spontaneity. Affective and perceptual biases in attention are examples of *automatic constraints* that serve to capture and hold one's attention on specific sources of information (see Section 1; Todd et al., 2012; Christoff et al., 2016; Irving, 2016; Irving and Thompson, this volume). Although, little is known about the relationship between automatic constraints and spontaneous thought, preliminary evidence from clinical literature on depression and anxiety implicates an important role of the brain's salience network (Seeley et al., 2007; McMenamin et al., 2014; Ordaz et al., 2016; reviewed in Christoff et al., 2016).

Future research would also benefit from developing new methods to covertly assess spontaneous thoughts without relying on self-report. Despite the usefulness of the introspective approaches discussed above, self-report assessments come with several limitations. Participants are sometimes unaware of their own mental activity, and most studies do not assess participants' subjective level of awareness. Participants may also interpret questions and use self-report scales in different ways, sometimes being sensitive to perceived experimenter expectations. Additionally, the act of requiring participants to introspect about their mental activity may interfere with the natural course of cognition, and bias participants to think in particular ways. To overcome some of these limitations, analysis methods such as machine learning algorithms could be applied in future work to covertly predict the nature of mental activity based on voxelwise patterns of brain activity (i.e. with *multivoxel pattern analysis;* Tusche et al., 2014; Kragel et al., 2016), and/or concurrent behavioral, occulometric, neurophysiological, or neuroendocrine measures. Additionally, these approaches could eventually be used in conjunction with real-time fMRI to train people to become more aware of their thinking patterns (e.g., McCraig et al., 2011; McDonald et al., in review; see also Garrison et al., 2013a; 2013b), and improve their ability to stay on-task, or engage in productive forms of spontaneous thought.

Finally, although neuroimaging and behavioral approaches can offer insight into the neural underpinnings of spontaneous thought, these approaches are correlational at best, and are unable to reveal whether patterns of brain activity play a causal role in the initiation or dynamics of spontaneous thought. Future research should therefore make use of methods such as transcranial magnetic stimulation (TMS) or transcranial direct current stimulation (TDCS) to transiently disrupt or enhance activity in certain brain regions (Axelrod et al., 2015), and neuropsychological studies should consider assessing if and how spontaneous thoughts become altered in patients with focal cortical or subcortical lesions. Towards this end, intracranial electroencephalography also marks a promising area of future research (reviewed in Selimbeyoglu and Parvizi, 2010; Fox et al., 2016).

An old sufi parable attributed to Mulla Nasrudin might serve as an analogy for the history of research on spontaneous thought. A police officer approaches a drunk man who's searching for something beneath a lamppost, "What are you looking for?" "My keys, Sir," the drunk man replies. The police officer helps to look for a few minutes. Finding nothing, the officer asks, "Are you sure you lost them under the lamppost?" "No," says the drunk, "I lost them in the park." "Then why are you searching here?!?" "Because there's a light." Like the drunk man, the field of psychology may have neglected spontaneous thought for over a century because it is shrouded in darkness. From Behaviourism through the Cognitive Revolution, the field looked for psychological processes under the light of experimental tasks. Methodological innovations in neuroscience and psychology moved our gaze a little further, but still we look only at those forms of "mind-wandering" that can be illuminated by their contents. Now it's time to break out the flashlights, to step into the darkness wherein lies the dynamics of spontaneous thought.

**Acknowledgments**

We extend our gratitude to Evan Thompson, Randy Buckner, Jonathan Smallwood, Matt Dixon, Rebecca Todd, Chandra Sripada, and Dasha Zabelina for helpful feedback and scholarly discussion on our teams' recently published manuscripts featured in the present chapter. A Brain and Behavioral Research Foundation Grant supported J.R.A.-H. while preparing this chapter. K.C. was supported by grants from the Natural Sciences and Engineering Research Council (NSERC) (RGPIN 327317–11) and the Canadian Institutes of Health Research (CIHR) (MOP-115197). Z.C.I. was supported by a Social Sciences and Humanities Research Council of Canada (SSHRC) postdoctoral fellowship, the Balzan Styles of Reasoning Project and a Templeton Integrated Philosophy and Self Control grant. R.N.S. was supported by an Alzheimer's Association grant (NIRG-14-320049). K.C.R.F. was supported by a Vanier Canada Graduate Scholarship.

**Figure 1. A Dynamic Model of Spontaneous Thought.** Spontaneous thought spans a conceptual space, inclusive of night dreaming, mind-wandering, and creative thought, that is relatively free from two kinds of constraints: 1) deliberate constraints (x-axis), and 2) automatic constraints (y-axis). According to this model, adapted and extended from Christoff and colleagues (2016), ruminative and obsessive thought are not truly spontaneous in nature due to strong bottom-up, "automatic" constraints bias their content. The dynamics of thought – the way thoughts unfold and flow over time – represent an important element of this model. As shown in the bottom left box, thoughts that are free from both kinds of constraints should transition relatively quickly and span different phenomenological content (represented by different colors). Conversely, excessively constrained thoughts should have longer durations with similar content (bottom right box).

**Figure 2. Methods to Assess the Neuroscience of Spontaneous Thought in Humans.** Left Panel: Many different first-person approaches are used to assess the nature of trait- or state-like thought patterns, although most existing studies do not differentiate between spontaneous and constrained forms of thinking. Right Panel: Neuroimaging, psychophysiological, and occulometric approaches are increasingly being employed to covertly assess the neurocognitive correlates of spontaneous thought. MRI = magnetic resonance imaging; EEG = electroencephalography; ERP = event-related potential; iEEG = intracranial EEG; tDCS = transcranial direct current stimulation; TMS: transcranial magnetic stimulation.

**Figure 3. Meta-Analytic Findings Reveal Neuroimaging Correlates of Task-Unrelated and/or Stimulus-Independent Thought.** A meta-analysis of 10 fMRI studies demonstrates that many regions within the brain's default network (outlined in blue, using 7-network parcellations from Yeo et al., 2010) and the frontoparietal control network (outlined in red) are reliably engaged across studies of task-unrelated and/or stimulus-independent thought. Regions within the default network include: medial prefrontal cortex (mPFC), posterior cingulate cortex (PCC), parahippocampal cortex (PHC, a part of the medial temporal subsystem, see Yeo et al., 2010), inferior parietal lobule (IPL), angular gyrus (AngG), superior temporal sulcus (STS), and ventral lateral PFC (vlPFC). Regions within the frontoparietal control network include: rostral lateral prefrontal cortex (rlPFC), dorsal anterior cingulate cortex (dACC), and precuneus (preC).

Regions spanning other networks include mid insula (mid-Ins), somatosensory cortex (SS), and temporal pole (TP, extending into the dorsal medial subsystem of the default network). Note that the fMRI studies included in the meta-analysis do not differentiate between spontaneous and constrained forms of thinking, so it is unclear which regions are involved in spontaneous thought, and which are involved in exerting constraints on those thoughts (see text).

**Figure 4. Neural Underpinnings of Night Dreaming and Creative Thought.** A) A meta-analysis of neuroimaging studies on REM sleep (a sleep stage characterized by dreaming) reveals greater activity in a number of brain regions compared to awake rest. Among others, these include medial temporal and medial prefrontal regions within the default network, and visual cortex. B) Creative thinking is associated with distinct temporal activity dynamics. The medial temporal lobe becomes engaged to a greater degree early in the creative process while generating a creative idea. Other regions within the default network, as well as key frontoparietal control network regions, become engaged to a greater degree during later stages of creative thinking, when evaluating creative ideas. Figures adapted from Fox et al., 2015 (A), and Ellamil et al., 2011 (B).

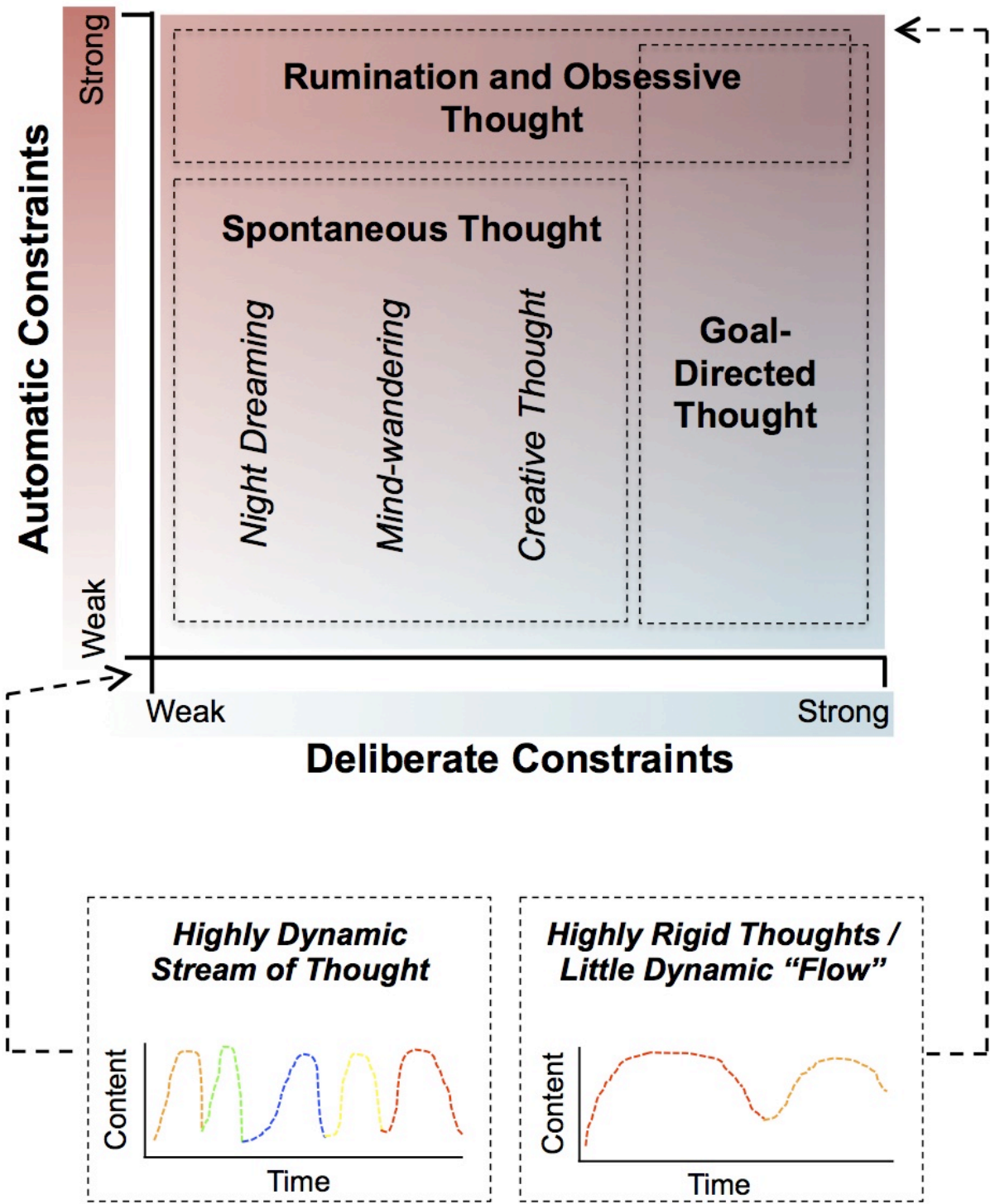

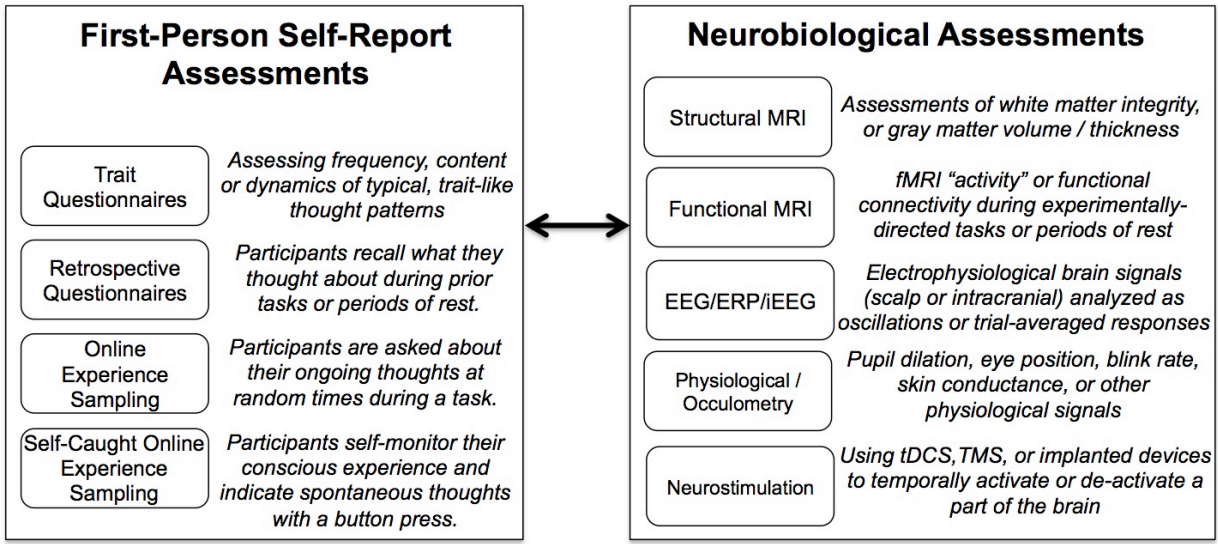

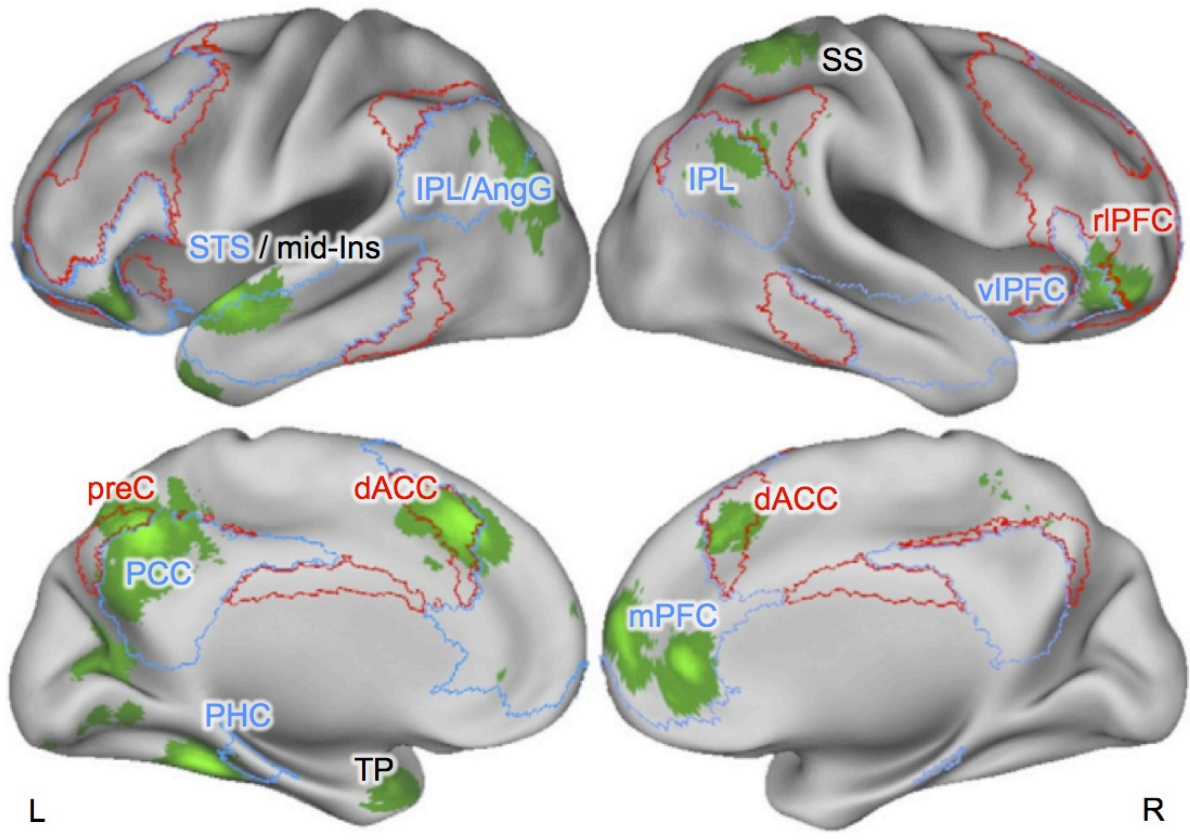

A.

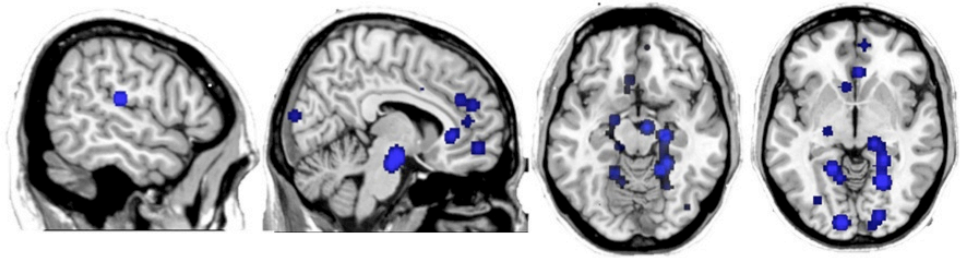

B.

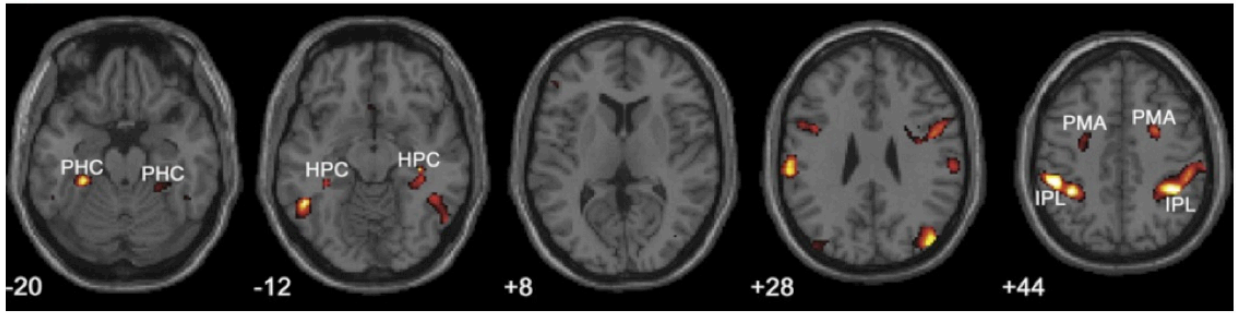

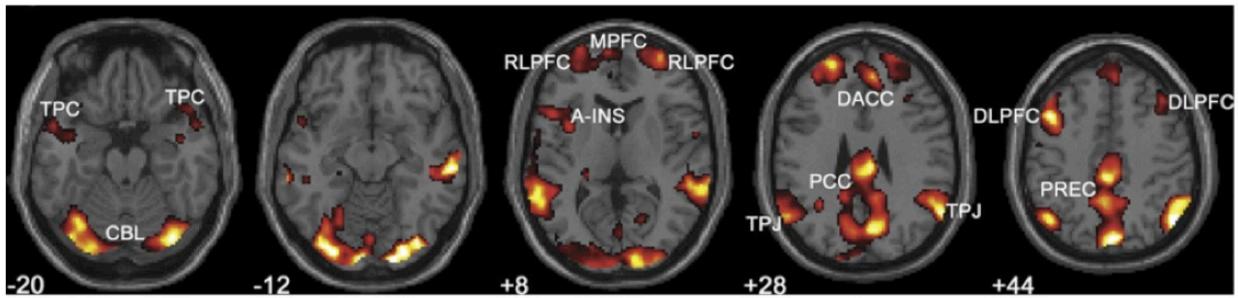

Hyatt, C. J., Calhoun, V. D., Pearlson, G. D., & Assaf, M. (2015). Specific default mode subnetworks support mentalizing as revealed through opposing network recruitment by social and semantic FMRI tasks. *Human Brain Mapping*, *36*, 3047–63.

Irish, M., & Piolino, P. (2016). Impaired capacity for prospection in the dementias--Theoretical and clinical implications. *The British Journal of Clinical Psychology / the British Psychological Society*, *55*, 49–68.

Irving, Z. C. (2016). Mind-wandering is unguided attention: accounting for the "purposeful" wanderer. *Philosophical Studies*, *173*, 547.

Irving, Z. C. & Thompson, E. (forthcoming). The Philosophy of Mind-Wandering. In K. C. R. Fox and K. Christoff (eds.) The Oxford Handbook of Spontaneous Thought: Mind-wandering, Creativity, Dreaming, and Clinical Conditions: New York: Oxford University Press.

Kaiser, R. H., Whitfield-Gabrieli, S., Dillon, D. G., Goer, F., Beltzer, M., Minkel, J., … Pizzagalli, D. a. (2015). Dynamic Resting-State Functional Connectivity in Major Depression. *Neuropsychopharmacology : Official Publication of the American College of Neuropsychopharmacology*, 1–9.

Kam, J. W. Y., Dao, E., Farley, J., Fitzpatrick, K., Smallwood, J., Schooler, J. W., & Handy, T. C. (2011). Slow Fluctuations in Attentional Control of Sensory Cortex. *Journal of Cognitive Neuroscience*, *23*, 460–470.

Kam, J. W. Y., Dao, E., Stanciulescu, M., Tildesley, H., & Handy, T. C. (2013). Mind wandering and the adaptive control of attentional resources. *Journal of Cognitive Neuroscience*, *25*, 952–60.

Kam, J. W. Y., & Handy, T. (forthcoming). Mind wandering and events in the external world: Electrophysiological evidence for attentional decoupling. In K. C. R. Fox and K. Christoff (eds.) The Oxford Handbook of Spontaneous Thought: Mind-wandering, Creativity, Dreaming, and Clinical Conditions: New York: Oxford University Press.

Kane, M. J., Brown, L. H., McVay, J. C., Silvia, P. J., Myin-Germeys, I., & Kwapil, T. R. (2007). For whom the mind wanders, and when: an experience-sampling study of working memory and executive control in daily life. *Psychological Science*, *18*, 614–21.

Kane, M. J., & McVay, J. C. (2012). What Mind Wandering Reveals About Executive-Control Abilities and Failures. *Current Directions in Psychological Science*, *21*, 348–354.

Killingsworth, M. A., & Gilbert, D. T. (2010). A wandering mind is an unhappy mind. *Science*, *330*, 932.
To appear in: *The Oxford Handbook of Spontaneous Thought* (Fox & Christoff, Eds.)     40

McDonald, A. et al., The Real-time fMRI Neurofeedback Based Stratification of Default Network Regulation Neuroimaging Data Repository. doi: http://dx.doi.org/10.1101/075275

McCaig, R. G., Dixon, M., Keramatian, K., Liu, I., & Christoff, K. (2011). Improved modulation of rostrolateral prefrontal cortex using real-time fMRI training and meta-cognitive awareness. *NeuroImage*, *55*, 1298–305.

McGuire, P. K., Paulesu, E., Frackowiak, R. S., & Frith, C. D. (1996). Brain activity during stimulus independent thought. *Neuroreport*, *7*, 2095–2099.

McKiernan, K. a, D'Angelo, B. R., Kaufman, J. N., & Binder, J. R. (2006). Interrupting the "stream of consciousness": an fMRI investigation. *NeuroImage*, *29*, 1185–91.

McMillan, R. L., Kaufman, S. B., & Singer, J. L. (2013). Ode to positive constructive daydreaming. *Frontiers in Psychology*, *4*, 626.

McVay, J. C., & Kane, M. J. (2010). Does mind wandering reflect executive function or executive failure? Comment on Smallwood and Schooler (2006) and Watkins (2008). *Psychological Bulletin*, *136*, 188–97; discussion 198–207.

Metzinger, T. (2013). The myth of cognitive agency: subpersonal thinking as a cyclically recurring loss of mental autonomy. *Frontiers in Psychology*, *4*, 931.

Mowlem, F. D., Skirrow, C., Reid, P., Maltezos, S., Nijjar, S. K., Merwood, A., … Asherson, P. (2016). Validation of the Mind Excessively Wandering Scale and the Relationship of Mind Wandering to Impairment in Adult ADHD. *Journal of Attention Disorders*. doi:10.1177/1087054716651927

O'Callaghan, C., Shine, J. M., Lewis, S. J. G., Andrews-Hanna, J. R., & Irish, M. (2015). Shaped by our thoughts--a new task to assess spontaneous cognition and its associated neural correlates in the default network. *Brain and Cognition*, *93*, 1–10.

O'Keefe, J., & Nadel, L. (1978). *The Hippocampus as a Cognitive Map* (pp. 1–570). Oxford, England: Oxford University Press.

O'Reilly, R. C. (2010). The What and How of prefrontal cortical organization. *Trends in Neurosciences*, *33*, 355–61.

Ordaz, S. J., Lemoult, J., Colich, N. L., Prasad, G., Pollak, M., Price, A., … Gotlib, I. H. (2016). Ruminative Brooding Is Associated with Salience Network Coherence in Early Pubertal Youth. *Social Cognitive and Affective Neuroscience*.

Patterson, K., Nestor, P. J., & Rogers, T. T. (2007). Where do you know what you know? The representation of semantic knowledge in the human brain. *Nature Reviews. Neuroscience*, *8*, 976–87.
To appear in: *The Oxford Handbook of Spontaneous Thought* (Fox & Christoff, Eds.)     42